\documentclass[aip,jcp,twocolumn,showpacs,superscriptaddress,amsmath,amssymb,floatfix,10pt,reprint]{revtex4-1} 
\usepackage{graphicx}  
\usepackage{dcolumn}   
\usepackage{bm}        
\usepackage{amssymb}   
\usepackage{framed}
\usepackage{amsmath}
\usepackage{hhline}
\usepackage{xcolor}
\usepackage{subfigure,amsmath,verbatim,moreverb}
\usepackage{tabularx}
\usepackage{lipsum}
\usepackage{longtable}
\usepackage{booktabs}
\usepackage{adjustbox}

\usepackage{etoolbox}
\AtBeginEnvironment{align}{\setcounter{subeqn}{0}}
\newcounter{subeqn} %

\begin{document}

\title{Efficient yet Accurate Dispersion-Corrected Semilocal Exchange-Correlation 
Functionals For Non-Covalent Interactions}

\author{Abhilash Patra}
\email{abhilashpatra@niser.ac.in}
\affiliation{School of Physical Sciences, National Institute of Science Education and Research, HBNI, Bhubaneswar 752050, India}
\author{Subrata Jana}
\email{subrata.jana@niser.ac.in, subrata.niser@gmail.com}
\affiliation{School of Physical Sciences, National Institute of Science Education and Research, HBNI, Bhubaneswar 752050, India}
\author{Lucian A. Constantin}
\email{lucian.constantin@iit.it}
\affiliation{Center for Biomolecular Nanotechnologies @UNILE, Istituto Italiano di Tecnologia, Via Barsanti, I-73010 Arnesano, Italy}
\author{Prasanjit Samal}
\email{psamal@niser.ac.in}
\affiliation{School of Physical Sciences, National Institute of Science Education and Research, HBNI, Bhubaneswar 752050, India}

\date{\today}

\begin{abstract}

Due to several attractive features, the meta-generalized-gradient approximations (meta-GGAs) are considered to be  
the most advanced and potentially accurate semilocal exchange-correlation functionals in the rungs of 
the Jacob's ladder of Density Functional Theory. So far, several meta-GGA are proposed by fitting to the test sets 
or/and satisfying as many as known exact constraints. Although the density overlap is treated by modern meta-GGA 
functionals efficiently, for non-covalent interactions, a long-range dispersion correction is essential. In this 
work, we assess the benchmark performance of different variants of the Tao-Mo semilocal functional (i.e. TM 
of Phys. Rev. Lett. {\bf 117}, 073001 (2016) and revTM of J. Phys. Chem. A {\bf 123}, 6356 (2019)) with 
Grimme's D3 correction for the several non-covalent interactions, including dispersion and hydrogen
bonded systems. We consider the zero, Becke-Johnson(BJ), and optimized power (OP) damping functions
within the D3 method, with both TM and revTM functionals. It is observed that the overall performance of the functionals 
gradually improved from zero to BJ and to OP damping. However, the constructed ``OP'' corrected (rev)TM+D3(OP) functionals perform
considerably better compared to other well-known dispersion corrected functionals. Based on the accuracy of the proposed functionals, 
the future applicability of these methods is also discussed.

\end{abstract}

\maketitle

\section{Introduction}
Semilocal exchange-correlation (XC) density functionals are the most preferred choice of doing electronic structure 
calculations within the Kohn-Sham (KS) Density Functional Theory (DFT) \cite{hohenberg1964inhomogeneous,kohn1965self}. 
Starting from the local density approximation (LDA)~\cite{kohn1965self,perdew1981self} to the higher rungs of the 
Jacob's ladder classification of XC functionals \cite{perdew2001jacob}, the semilocal approximations are characterized 
as the generalized gradient approximations (GGAs) \cite{scuseriaREVIEW05,becke1988density,lee1988development,
perdew1992atoms,perdew1996generalized,armiento2005functional,constantin2009exchange,
wu2006more,zhao2008construction,constantin2010communication,
constantin2015gradient,constantin2011correlation,fabiano2014generalized,constantin2016semiclassical,cancio2018fitting,
constantin2008dimensional,pvrecechtvelovaJCP14,pvrecechtvelovaJCP15,henderson2008generalized,carmona2018generalized,
peverati2012exchange} and meta-GGAs \cite{della2016kinetic,becke1989exchange,van1998novel,zhao2006new,perdew1999accurate,
tao2003climbing,constantin2012semilocal,sun2015semilocal,MN15-L,MN15,ruzsinszky2012meta,constantin2016semilocal,
sun2015strongly,tao2016accurate,wang2017revised,RM06-L,RM06,mezei2018simple,revTM2019,patra2019efficient,patra2019relevance,
patra2019laplacian,aschebrock2019ultranonlocality,sun2013density,RM11}. Higher rungs than meta-GGA use non-local information
from KS orbitals and eigenvalues, and are recognized from the point of view of their sophistication, as the
so-called rung 3.5 \cite{janesko2012nonspherical, janesko2013rung,janesko2010rung,janesko2012nonempirical,
janesko2018long,constantin2016hartree,constantin2017modified}, hybrids and hyper-GGA functionals
\cite{perdew2008density,perdew2005prescription,odashima2009nonempirical,arbuznikov2011advances,
jaramillo2003local,kummel2008orbital,becke2005real,becke2007unified,becke2003real,becke2013density,fabiano2014global,
sun2013semilocal,fabiano2013testing,jana2019long,doi:10.1063/1.2409292,C8CP00333E,C8CP00717A,C8CP06715E,JANA20181,
jana2018efficient}, double hybrids and DFT coupled-cluster based methods \cite{goerigk2010efficient,goerigk2017look,
smigaOEPh,bartlett2005exchange,bartlett2005ab,grabowski2013optimized, grabowski2014orbital}, adiabatic-connection 
methods and generalizations of the random phase approximation (RPA) \cite{seidl2000simulation,liu2009adiabatic,
sun2009extension,gori2009electronic,mirtschink2012energy, vuckovic2016exchange,fabiano2016interaction,
giarrusso2018assessment,fabiano2018investigation, ConstantinMISI2019,dobson2002correlation,constantin2007simple,
terentjev2018gradient,constantin2016simple,toulouse2005simple,richardson1994dynamical,
bates2016nonlocal,bates2017convergence,ruzsinszky2016kernel,dobson2000energy,
erhard2016power,patrick2015adiabatic}.

Meta-GGA XC functionals improve the overall performance of GGAs, and the hybrid methods do the same 
over their bare semilocal counterparts. But none of these functionals able to incorporate the long-range correlation, 
which is essential for systems dominated by weak bonds. For the last couple of decades, the formulation of meta-GGA
functionals has been made very physical insightful through the inclusion of short- and intermediate-range behavior 
of the weakly bonded systems \cite{sun2015strongly,zhao2006new}. However, studies show that semilocal approximations
do not incorporate short- and intermediate-range dispersion~\cite{PhysRevLett.121.113402,PhysRevLett.122.213001}. 
Designing density functionals, irrespective of the short- and intermediate-range dispersion or van der Waals (vdW) 
interactions as well as to retain their accuracy for the density overlap region, a long-range vdW correction is always 
necessary to describe the functional performance correctly for the binding energies of weakly 
bonded systems~\cite{goerigk2017look,GordonB97MV2016,GordonB97XV2014,GordonLCh2008,SCAND32016}. 

The long-range vdW interaction can be captured via~\cite{GrimmeJCC2006}
\begin{equation}
    E_{vdW} = -\sum_{i<j}\sum_{n=6,8,10,...}s_{n}\frac{C_{n,ij}}{r_{ij}^n}f_{dmp,n} (r_{ij} )~,
    \label{eq1}
\end{equation}
where the dispersion coefficients, $C_{n,ij}$ are determined either experimentally~\cite{GrimmeJCC2006} or 
theoretically
\cite{GrimmeJCC2006} and those may be chemically insensitive
; $s_n$ are density functional 
dependent global scaling parameters; and $r_{ij}$ is the inter-nuclear distance between the $i^{th}$ and $j^{th}$ 
atoms. This simple term, when added to any semilocal or hybrid density functionals leads to the well known dispersion 
corrected density functional (DFT+D) method~\cite{YangDFTD2002,GrimmeJCC2004}. To avoid the singularities at small $r_{ij}$ 
a damping function $f_{damp,n} (r_{ij} )$ is used. The form of the damping function plays a dramatic role in the 
functional performance when applied to the dispersion bonded systems. However, the cutoff procedure of the $vdW$ range
through the damping function should be judiciously chosen with caution avoid deterioration of the functional
performance or over-binding problem in case of non-covalent interaction. This is crucial for the H-bonded 
systems for which the inclusion of the dispersion correction deteriorates the functional performance. We will 
discuss all these points in our results section. 

Note that by truncating Eq. (\ref{eq1}) up to $n=6$ and choosing the damping function as,
\begin{equation}
    f^{D2}_{dmp,6}(r_{ij}) = \frac{1}{1+e^{-\alpha(r_{ij}/r_{0,ij}-1)}}~, 
    \label{eq2}
\end{equation}
one ends up with the Grimme's DFT-D2 model~\cite{GrimmeJCC2006}, where $r_{0,ij}$ is the sum of atomic $vdW$ radii, 
and $\alpha=20$ is chosen for better damping or steepness of this function~\cite{YangDFTD2002,GrimmeJCC2004}. 
The scaling factor $s_6$ depends on the choice of the particular density functional, and the $C_6$ coefficients are 
obtained by fitting to the binding energies ($\Delta E$) and inter-molecular distances of experimental or accurate 
theoretical values \cite{GrimmeJCC2004}. 
  
Grimme's DFT-D3 dispersion correction~\cite{GrimmeJCP2010} was proposed using both the $C_{6,ij}$, and $C_{8,ij}$ terms. 
However, the higher-order terms  corresponding to $n>6$ are more short-ranged and strongly influence the short-range 
part of the dispersion interaction~\cite{GrimmeJCP2010}. Several choices of the damping function are proposed,
improving the functional performance in different prospects. Among the different choices of damping 
functions, the widely used ones are: \\

(i) D3(0): The zero damping function is having the following analytic form~\cite{GrimmeJCP2010}:
\begin{equation}
   f^{D3(0)}_{dmp,n}(r_{ij}) = [1+6(\frac{r_{ij}}{s_{r,n}r_{0,ij}})^{-\alpha_n}]^{-1}~.  
\label{eq3}
\end{equation}
The adjustable parameters of $vdW$ energy terms and damping function are chosen as following: $s_{6} = 1$, 
$\alpha_6 = 14$, $\alpha_8 = 16$. Also, $s_{r,8}=1$ is chosen for most of the density functionals, leaving 
the parameters $s_{r,6}$ and $s_8$, that depend on the density functional form.\\

(ii) D3(BJ): The Becke-Johnson (BJ) damping function is having the 
form~\cite{BJD312005,BJD322005,BJD332006,BJD342005,GrimmeJCC2011}:
\begin{equation}
 f^{D3(BJ)}_{dmp,n}(r_{ij}) = \frac{r^n_{ij}}{r_{ij}^{n}+(\alpha_1r_{0,ij}+\alpha_2)^n}~.   
 \label{eq4}
\end{equation}
The rationale behind the chosen damping function form of the BJ is due to the exchange-hole dipole-moment (XDM) 
correction of Becke and Johnson~
\cite{BJD322005,BJD342005,BJD312005,JOHNSON2006333}. The BJ damping approaches to a constant value at small 
inter-atomic separation ($r_{ij}\to 0$), that differs from the D3(0). For most of the functionals, $s_{6}$ is 
generally fixed to unity.\\

(iii) D3(CSO): The Becke-Johnson damping function became the most preferred method for the DFT+D functional as it 
outperforms the D3(0) in most cases. However, later on, Schr\"{o}der et al.~\cite{CSOD32015} 
simplified it 
by proposing the $C_6$-Only (CSO) approach, where the eighth-order term is approximated within the sigmoidal
interpolation function. The damping function of D3(CSO) is given by~\cite{CSOD32015},
\begin{eqnarray}
 f^{D3(CSO)}_{dmp,6}(r_{ij}) &=& \frac{r^6_{ij}}{r_{ij}^6+(\alpha_3r_{0,{ij}}+\alpha_4)^6}\nonumber\\
 &&[1+ \frac{\alpha_1}{s_6[1+exp(r_{ij}-\alpha_2r_{0,ij})]}]~.    
 \label{eq5}
\end{eqnarray}
A closer look shows the similarities between Eqs.~(\ref{eq4}) and (\ref{eq5}). For most density 
functionals, 
Schr\"{o}der et al.~\cite{CSOD32015} fixed $\alpha_3 \approx 0$, $\alpha_4 \approx 6.25$, and 
$\alpha_2 \approx 2.5$. 
\\

(iv) The more general form of BJ damping is proposed recently by Witte et al.~\cite{GordonD3OP2017}. This is known 
as ``optimize-power" damping with the following analytic form 
\begin{equation}
    f^{D3(OP)}_{dmp,n}(r_{ij}) =\frac{r_{ij}^{\beta_n}}{r_{ij}^{\beta_n}+(\alpha_1r_{0,ij}+\alpha_2)^{\beta_n}}~.     
\label{eq6}
\end{equation}
The similarities between D3(BJ) and D3(OP) are noticeable. Most importantly, the parameter $\beta$ controls the rate 
of dispersion interaction. Here, $\beta_8 = \beta_6 + 2$, and the same dispersion coefficients and $vdW$ radii are 
used in D3(OP) damping. Also, similar to the D3(BJ), the parameters $\alpha_1$ and $\alpha_2$ control the distance 
where the damping function corresponding to the dispersion correction will be switched on or off. It was also shown 
that the D3(OP) improves the descriptions of weakly bonded molecular systems when coupled with any density 
functionals \cite{GordonD3OP2017}. 

These dispersion correction methods are important for our present study. There are several studies on the 
performance of the density functionals with dispersion corrections~\cite{GrimmeJCTC2010,goerigk2010efficient,GrimmePCCP2011,goerigk2017look,
GordonB97MV2016,GordonB97XV2014,GordonLCh2008, JohnsonJPOC2009,PerdewJCTC2013,GrimmePCCP2006,KronikJCTC2011}. 
Several recent, accurate meta-GGA density functionals suitable for quantum chemical calculations are also proposed 
and tested for a broad range of systems~\cite{SCAND32016,TaoIJMPB2019}. However, these 
functionals are not benchmarked for a wide range of molecular properties. The motivation of the present study 
follows from the very accurate performance of the different variants of TM semilocal functionals (TM 
\cite{tao2016accurate} and revTM\cite{revTM2019}) for quantum chemistry. Here, we combine the D3(0), D3(BJ), 
and D3(OP) with the TM and revTM functionals to assess their performance for non-covalent interaction test sets and 
H bonded water systems. We observe that the combination of the TM and revTM with D3(OP) gives improvements over various other 
combinations proposed so far. Most importantly, the TM+D3(OP) and revTM+D3(OP) do not deteriorate much the H-bond 
energies compared to their base functional accuracy. To present the functionals performance, we arrange our paper as 
follows: In the following, we briefly review the TM and revTM meta-GGA functionals, and we construct their dispersion 
corrected terms. Next, we test the proposed functionals concerning different non-covalent interaction test sets. Lastly, 
we conclude and summarize our results based on insightful analysis.

\section{Theory}

\subsection{Review of TM and revTM functionals}

The construction of TM and revTM functionals have already been reviewed in Ref.~\cite{revTM2019}. Here, we only focus 
on the key differences between these two functionals, as outlined below. 

\begin{figure}[t]
\begin{center}
\includegraphics[width=7.0cm,height=5.0cm]{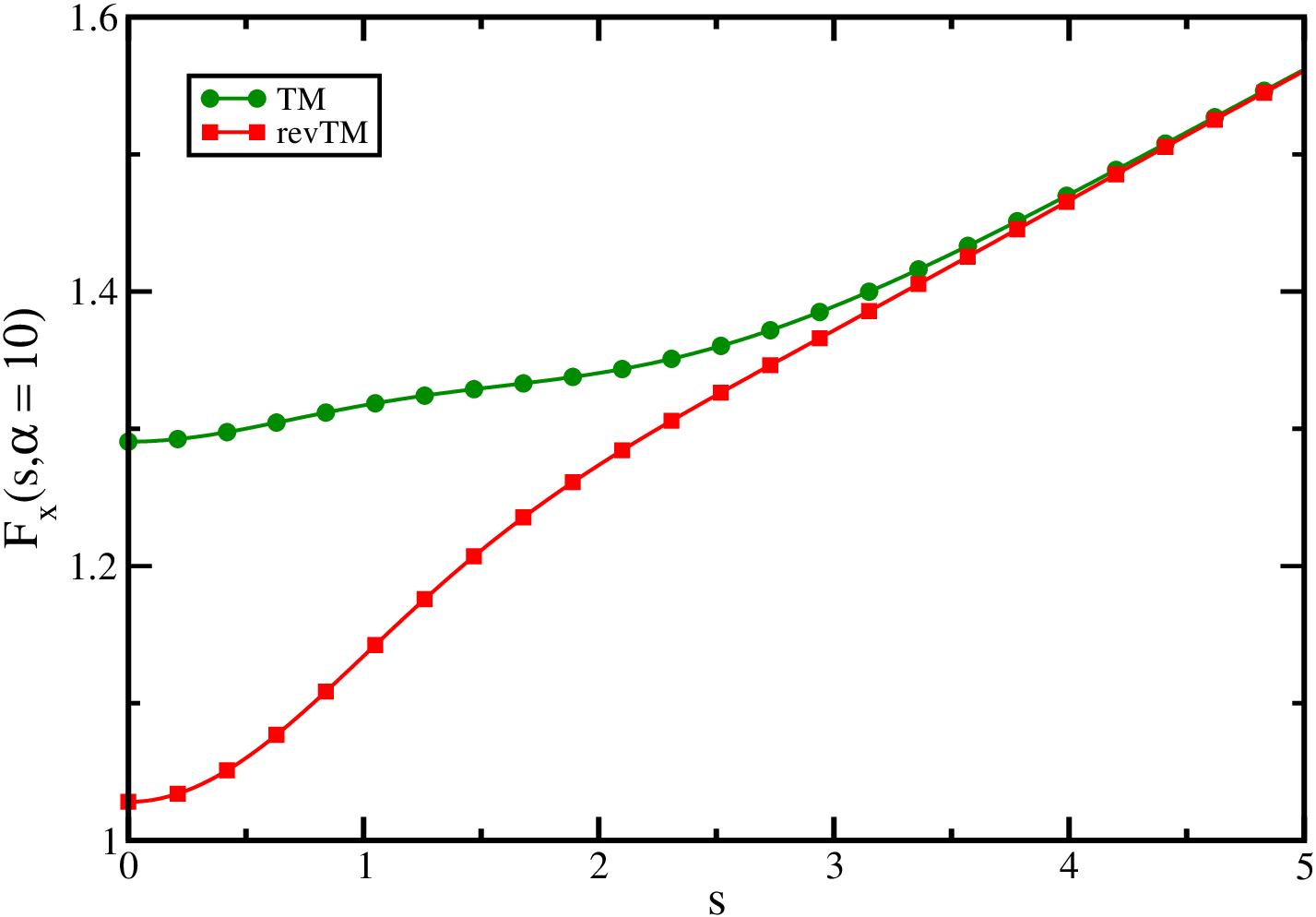}\\
\vspace{0.5 cm}
\includegraphics[width=7.0cm,height=5.0cm]{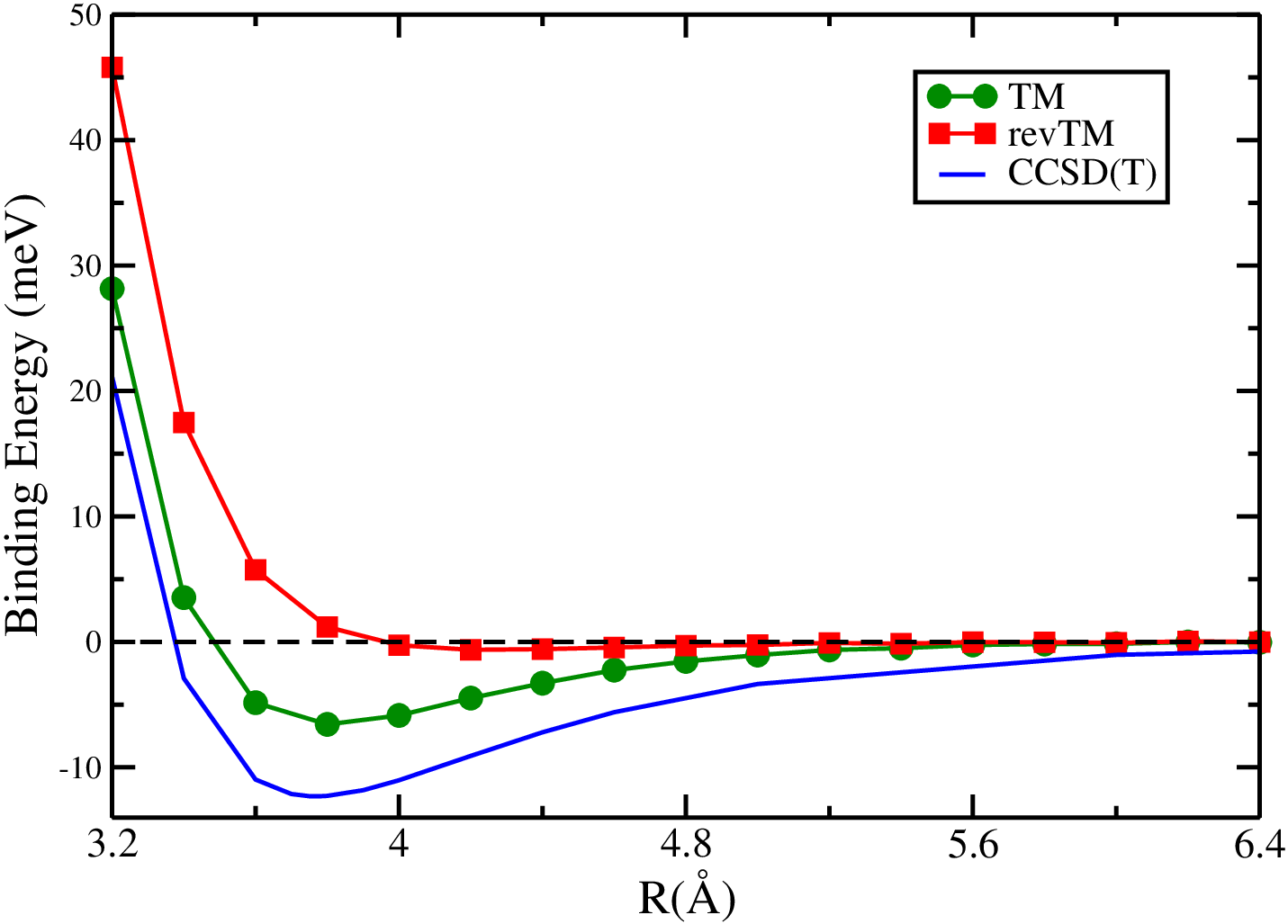}
\end{center}
\caption{(Upper panel) Exchange enhancement factors of TM and revTM functionals. (Lower panel) The binding energy of
Ar$_2$ dimer obtained from TM and revTM functionals. For XC integrals the 99 points radial grid and 590 points angular 
Lebedev grid are used}
\label{exen}
\end{figure}
(i) Firstly, the significant difference between the TM and revTM exchange functionals comes from
modeling the reduced Laplacian $\tilde{q}$ in the slowly varying (sc)
fourth-order gradient approximation (GE4) of the exchange enhancement factor.  
Thus, the revTM uses $\tilde{q}_b = \frac{9(\alpha-1)}{20[1+b\alpha (\alpha-1)]^{1/2}}+\frac{2p}{3}$ (with $b=0.40$) 
instead of $\tilde{q} = \frac{9}{20}(\alpha-1)+\frac{2p}{3}$ as found in the TM functional.
As a result, in the overlapping closed shells~\cite{sun2013density} ($\alpha>>1$, $s\approx 0$) $F_x^{TM}$ and 
$F_x^{revTM}$ differ from each other drastically. This is shown in Fig.~\ref{exen}, where, in the upper panel, we 
have plotted the exchange enhancement factors of both functionals for $\alpha=10$.  Note that this modification 
affects the lattice constants of the alkali metals, ionic solids and layered materials~\cite{revTM2019,patra2019performance}.

The TM functional is very accurate for several solid-state and molecular properties
~\cite{mo2017performace,mo2017assessment,mo2018aip,TangMRE2018,tian2017computation,mo2016performance}. 
Specially, the best performance of the TM functional is evident from the lattice constants of the ionic solids~\cite{TangMRE2018,mo2017assessment} 
and hydrogen-bonded complexes~\cite{mo2016performance,mo2017performace}.
In refs.~\cite{mo2016performance,mo2017assessment,TangMRE2018,tian2017computation} it has been argued that the TM exchange enhancement factor
shares  slightly oscillatory behavior to some extent as it is shown in Voorhis-Scuseria (VSXC) ~\cite{van1998novel} and M06-L~\cite{zhao2006new} 
functionals~\cite{johnson2009oscillation}. 
On the other-hand the revTM exchange enhancement factor behaves differently in the region $\alpha >> 1$ and $s\approx 0$ which is
important for the overlapping of the closed shells or weekly bonded systems. 
But none of the functionals (including TM and revTM) do not incorporate correct 
$1/r_{ij}^6$ form or the long-range interaction or correct dispersion physics~\cite{johnson2009oscillation}. 

To further elaborate on this point, and distinguish the different behavior of the TM and revTM for weekly interacting systems, we 
also plot the binding energy curve of Ar$_2$ dimer for both the functionals (shown in the lower panel of Fig.~\ref{exen}). From the figure, 
we observe that the bare revTM functional 
unbound the Ar$_2$ dimer because $\tilde{q}_b<\tilde{q}$ in the middle of the bonding region. 
The behavior of the $\alpha$ and $\tilde{q}_b$ can be found in Fig. 4 and Fig. 5 of the ref.~\cite{della2016kinetic}.
The difference in capturing the interaction by 
both the functionals are important for non-covalent bonded molecules. Note
that bare TM functional is already quite good without including any $vdW$ correction. 

The behavior of the TM and revTM functionals can also be understood from the recent
investigation of the functionals performance for the water clusters~\cite{jana2020accurate}. In 
ref.~\cite{jana2020accurate} it is shown that both the TM and revTM predicts correctly the ordering stability 
of the water hexamers, whereas, the revTM is quite good for overall performance of water and ice structures.   

(ii) Secondly, the other important difference is arising due to the correlation content of both the functionals. In 
revTM, the linear response parameter $\beta$ has been generalized to the form of the exact, density-dependent 
second-order gradient expansion (GE2) parameter proposed in the revTPSS meta-GGA \cite{PerdewPRL2009}
correlation energy functional. We recall that TM correlation functional uses the high-density GE2 parameter (a.i. 
$\beta=0.066725$). The revTM also keeps all the useful features of the TM correlation by making the 
correlation energy functional spin-independent in the low-density or strong-interaction limit~\cite{tao2016accurate}. The change 
in correlation energy functional improves the jellium surface XC energies \cite{revTM2019}, which are relevant for 
the surface energies of simple metals. Note that the change in the correlation does not affect the non-covalent 
interaction systems. 

\begin{figure}
 \begin{center}
 \includegraphics[width=3.2in,height=2.4in,angle=0.0]{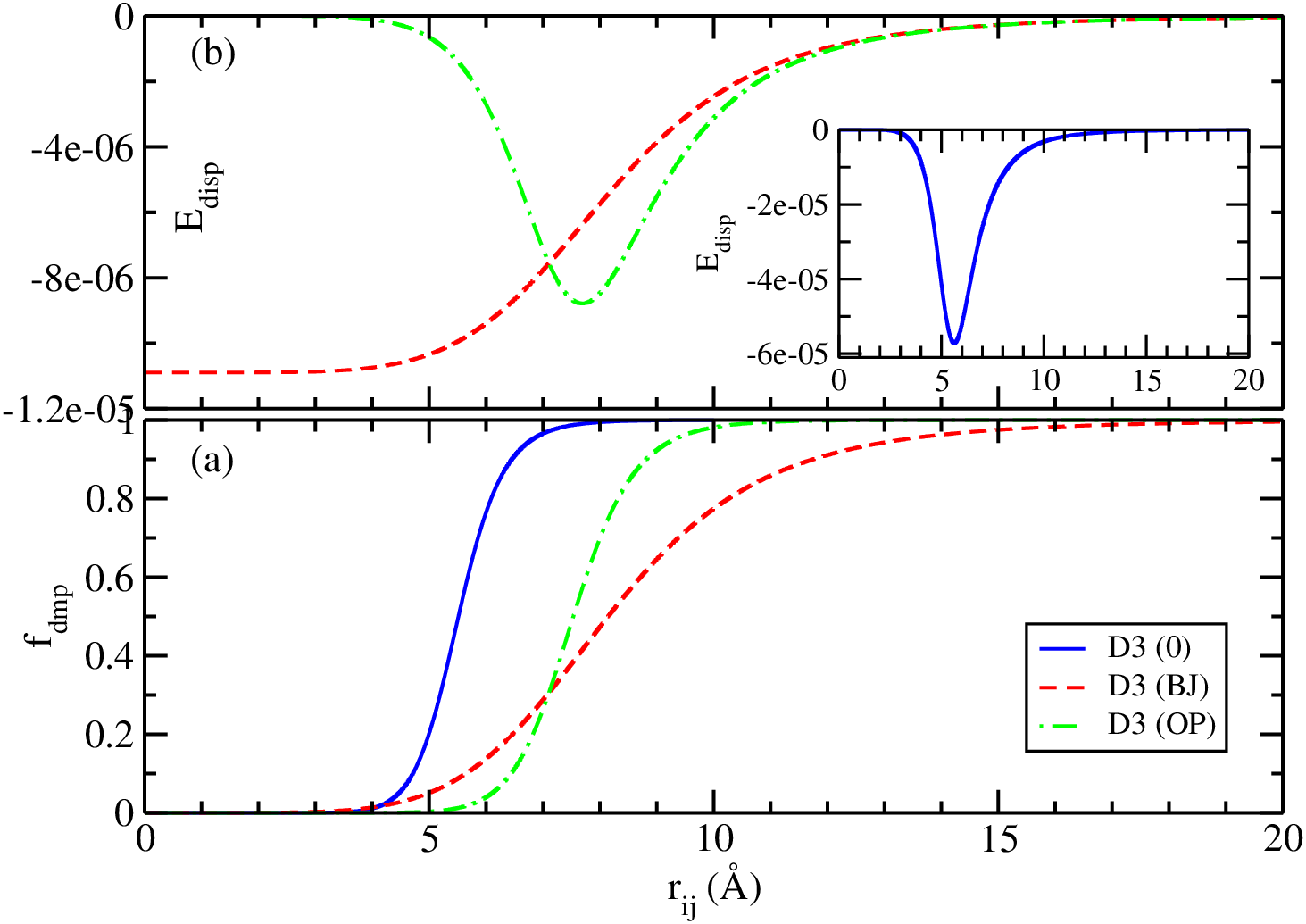}
  \end{center}
  \caption{The lower panel shows the behavior of damping functions with respect to inter-nuclear distance $r_{ij}$.
   The upper panel shows the contribution to the dispersion energy of Ne$_2$ for three types of damping functions.
   For all the figures, we have used $r_{0,ij}=3.3$~\AA, and for Ne$_2$, $C_6$=6.35
   . The inset of the upper panel is for D3(0), as the scale of energy is different.}
  \label{fig_damp}
\end{figure}
\begin{table}
\caption{Parameters used in the calculations.}
\begin{tabular}{ccccccccccccccccccccccccccccccc}\hline\hline
Parameters&	TM&	revTM&	\\ \hline
\multicolumn{2}{c}{\bf D3(0)(Zero damping)}\\
s$_6$	&	1.00&	1.00&	\\
s$_8$	&	0.00&	0.00&	\\
s$_{r,6}$	&	1.47&	1.16&	\\
\multicolumn{2}{c}{\bf D3(BJ)(Becke and Johnson damping)}\\
s$_6$	&	1.00&	1.00&	\\
s$_8$	&	0.00&	0.00&	\\
a$_1$	&	0.825&	0.23&	\\
a$_2$	&	5.42&	5.42&	\\
\multicolumn{2}{c}{\bf D3(OP)(Optimized Power damping)}\\
s$_6$	&	1.00&	1.00&	\\	
s$_8$	&	0.00&	0.00&	\\
$\alpha_1$&	0.64&	0.24&	\\
$\alpha_2$&	5.42&	5.42&	\\
$\beta$	&	14.00&	14.00&	\\ \hline \hline
\end{tabular}
\label{s22_param}
\end{table}

\subsection{Dispersion corrected TM and revTM functionals}

To construct the dispersion corrected functionals, we combine D3(0), D3(BJ), and D3(OP) dispersion 
corrections with the TM and revTM functionals. To determine the dispersion parameters associated with
the functionals, one needs to fit the functional with appropriate non-covalent interaction test set. 
The most preferred choice is to use the S22 test set of Jurecka et. al.~\cite{S22test2006}. This test 
set is chosen wisely as it consists of hydrogen bonded complexes, dispersion bonded complexes, and mixed 
complexes (having both interaction types). In our calculations, we consider the new benchmark CCSD(T) values
of Marshall et. al.~\cite{MarshallJCP2011}, along with the geometries available from GMTKN55 test 
set~\cite{goerigk2017look}. The optimized parameters of the respective functionals are summarized 
in Table~\ref{s22_param}. For all the functionals, we consider the standard $s_{6}=1$, and we obtain
$s_{8}=0$, as any other value of $s_{8}$ increases the mean absolute error (MAE) of the S22 test set. 
Also, the SCAN+D3(0) has been also proposed by considering $s_{8}=0$. Though the revTM functional proposed from
the TM functional, we do not observe any improvement in the error statistics by incorporating 
the $s_{8}$-term in the revTM+D3 functionals. Therefore, we keep only the $s_{6}$-term, and the dispersion 
parameters are fixed by minimizing the MAE of the S22 test set. Note that the revTM functional demands more
dispersion interaction than the TM functional due to its more unbound nature for dispersion bonded systems.

To understand the role played by and impact of different parameters on the damping function as well as energy 
component, in Fig.~\ref{fig_damp}, we plot the damping function (lower panel) and $E_{vdW}$ of Eq. (\ref{eq1}) 
(upper panel), in case of the Ne$_2$ dimer for which the $C_6$ coefficient 
is known. By construction, the DFT+D3(BJ) damping approach shows constant value at small inter-atomic separation, 
while the D3(OP) works within D3(0) and D3(BJ).    

It is noteworthy to mention that in this work the 3-body term is used with all the D3 schemes, being 
\cite{AxilrodJCP1943},
\begin{equation}
  E^{D3}_{3-body}=-\frac{1}{6}\sum_{A,B,C}^{triples}\frac{C_9^{ABC}(1+3\cos\phi_A\cos\phi_B\cos\phi_C)}
{r^9_{ABC}}
 \times f_9^d(r_{ABC})~,
\label{eq7}
\end{equation}
where the damping function $f_9^d$ is related to the D$3$ dispersion interaction coupled with the correlation part 
of the semilocal density functional. Here $\phi_A$, $\phi_B$ and $\phi_C$ are the angles formed of by the three 
atoms $A$, $B$ and $C$, and $r_{ABC}$ is the geometric mean distance. We recall that the 3-body term represents only 
a small fraction of the total dispersion interaction, being analyzed in several works ~\cite{DiStasio_2014,
DobsonIJQC2014,SherrillJCP2014}.

\begingroup
\begin{table}
\caption{Tabulated are the test sets used in our present calculations. All geometries are taken from 
Ref. \cite{goerigk2017look}, with exception of the L7, DSCONF, and MG8 test sets, where 
we used the geometries from  the respective reference articles.}
\begin{adjustbox}{max width=0.5\textwidth}
\begin{tabular}{lccccccccccccccccccccccccccccccc}\hline\hline
Test Set&Description\\	 \hline
S22&22 non-covalent interactive complexes~\cite{S22test2006,MarshallJCP2011}\\
L7&7 large molecular binding energies~\cite{L7JCTC2013}\\
S66&66 non-covalent interactive complexes~\cite{S66JCTC2011}\\
ADIM6&6 n-alkane dimers interaction energies~\cite{GrimmeJCP2010,goerigk2017look}\\
AHB21&21 neutral anion dimers interaction 
energies~\cite{LaoJCTC2015,KniziaJCP2009,GordonCPL1989}\\
CARBHB12&12 hydrogen-bonded complexes~\cite{goerigk2017look}\\
CHB6&6 cation-neutral dimers interaction energies~\cite{LaoJCTC2015,KniziaJCP2009,GordonCPL1989}\\
HAL59&59 halogenated dimers interaction energies~\cite{HAL59JCTC2012,HAL59JCTC2013}\\
HEAVY28&28  heavy element hydrides interaction energies~\cite{GrimmeJCP2010,goerigk2017look}\\
IL16&6 anion-cation dimers interaction energies~\cite{LaoJCTC2015,KniziaJCP2009,GordonCPL1989}\\
PNICO23&23 pnicogen-containing dimers interaction 
energies~\cite{KartonJCP2012,PnicogenJPCA2015,goerigk2017look}\\
RG18&18 rare-gas complexes interaction 
energies~\cite{RG18MP1970,RG18JCC2013}\\
ACONF& alkane conformers interaction energies~\cite{ACONFJPCA2009}\\
Amino20$\times$4&20 amino acid conformers interaction 
energies~\cite{AMINO20JCTC2016,AMINO20JCTC2009}\\
BUT14DIOL&14 butane-1,4-diol conformers interaction energies~\cite{BUT14JPCA2013,goerigk2017look}\\
ICONF&inorganic systems~\cite{goerigk2017look} \\
IDISP&Intramolecular dispersion interactions~\cite{GrimmeACIE2006,GrimmePCCP2007, 
goerigk2010efficient,GrimmeJOC2007}\\
MCONF&Melatonin conformers interaction energies~\cite{MCONFJPCA2013,goerigk2017look}\\
PCONF21&tri and tetrapeptide conformers~\cite{PCONF2013,GrimmeJCTC2010,PCONF2005}\\
SCONF&sugar conformers~\cite{GrimmeJCTC2010,SCONF2009} \\
UPU23&RNA-backbone conformers~\cite{UPU23JCTC2015}\\
WATER27&27 charged/neutral water clusters binding energies~\cite{WATER27JCTC2009,WATER27JCTC2017}\\
DSCONF&30 conformers of Lactose, Maltose, and Sucrose~\cite{chan2020aqueous}\\
MG8& 64 small representative thermochemical test set~\cite{chan2018formulation}\\ 
%
\hline\hline
\label{test_set}
\end{tabular}
\end{adjustbox}
\end{table}
\endgroup

\begin{table*}
\caption{Interaction energies (in kcal/mol) of S22 data set. The mean error (ME) and mean absolute error (MAE) are 
also reported. The best values are marked with bold style.}
\label{S22-full}
\begin{adjustbox}{max width=\textwidth}
\begin{tabular}{lllccccccccccccccccccccccccccccccc}\hline\hline
S22 complex			&	CCSD(T)	&	TM+D3(0)	&TM+D3(BJ)&	TM+D3(OP)&	revTM+D3(0)&	revTM+D3(BJ)&	revTM+D3(OP)&	\\ \hline
\multicolumn{8}{c}{\bf Hydrogen bonded complexes}													\\
NH$_3$ dimer ($C_{2h}$ )	&	3.133	&	3.397	&	3.403	&	3.352	&	3.257	&	3.303	&	3.150	&	\\
H$_2$O dimer ($C_s$ )		&	4.989	&	5.360	&	5.361	&	5.317	&	5.373	&	5.394	&	5.247	&	\\
Formic acid dimer ($C_{2h}$ )	&	18.753	&	18.771	&	18.852	&	18.717	&	19.033	&	19.258	&	18.801	&	\\
Formamide dimer ($C_{2h}$ )	&	16.062	&	15.740	&	15.782	&	15.658	&	15.968	&	16.139	&	15.791	&	\\
Uracil dimer ($C_{2h}$ )	&	20.641	&	19.708	&	19.677	&	19.623	&	20.068	&	20.211	&	19.890	&	\\
2-pyridone-2-aminopyridine ($C_1$ )&	16.934	&	16.608	&	16.583	&	16.521	&	17.134	&	17.250	&	16.922	&	\\
Adenine-thymine WC ($C_1$ )	&	16.660	&	15.931	&	15.905	&	15.840	&	16.395	&	16.462	&	16.132	&	\\ [0.1 cm]
ME				&	$-$	&	-0.23	&	-0.22	&	-0.30	&{\bf	0.00}	&	0.12	&	-0.17	&	\\
MAE				&	$-$	&	0.42	&	0.44	&	0.46	&	0.27	&	0.30	&{\bf	0.26}	&	\\ [0.1 cm]
\multicolumn{8}{c}{\bf Dispersion bonded complexes}													\\
CH$_4$ dimer (D$_{3d}$ )	&	0.527	&	0.578	&	0.568	&	0.517	&	0.635	&	0.470	&	0.479	&	\\
C$_2$H$_4$ dimer (D$_{2d}$ )	&	1.472	&	1.633	&	1.691	&	1.599	&	1.418	&	1.453	&	1.269	&	\\
Benzene-CH$_4$ (C$_3$ )		&	1.448	&	1.560	&	1.482	&	1.460	&	1.470	&	1.459	&	1.413	&	\\
Benzene dimer (C$_{2h}$ )	&	2.654	&	2.670	&	2.663	&	2.531	&	2.772	&	2.908	&	3.267	&	\\
Pyrazine dimer (C$_s$ )		&	4.255	&	3.999	&	4.076	&	3.847	&	4.018	&	4.196	&	4.317	&	\\
Uracil dimer (C$_2$ )		&	9.805	&	10.028	&	10.028	&	9.730	&	9.756	&	9.704	&	9.700	&	\\
Indole-benzene (C$_1$ )		&	4.524	&	4.432	&	4.467	&	4.267	&	4.406	&	4.723	&	5.078	&	\\
Adenine-thymine (C$_1$ )	&	11.730	&	12.127	&	12.179	&	11.792	&	11.397	&	11.644	&	11.524	&	\\ [0.1 cm]
ME				&	$-$	&	0.07	&	0.09	&	-0.08	&	-0.06	&{\bf	0.01}	&	0.07	&	\\
MAE				&	$-$	&	0.16	&	0.15	&	0.13	&	0.12	&{\bf	0.09}	&	0.22	&	\\ [0.1 cm]
\multicolumn{8}{c}{\bf Hydrogen + dispersion (mixed) bonded complexes}											\\
C$_2$H$_4$-C$_2$H$_2$ (C$_{2\nu}$ )&	1.496	&	1.486	&	1.486	&	1.458	&	1.598	&	1.588	&	1.537	&	\\	
Benzene-H$_2$O (C$_s$ )		&	3.275	&	3.886	&	3.863	&	3.754	&	3.874	&	3.781	&	3.646	&	\\	
Benzene-NH$_3$ (C$_s$ )		&	2.312	&	2.600	&	2.543	&	2.480	&	2.564	&	2.498	&	2.427	&	\\	
Benzene-HCN (C$_s$ )		&	4.541	&	4.657	&	4.685	&	4.597	&	4.420	&	4.741	&	4.531	&	\\	
Benzene dimer (C$_{2\nu}$ )	&	2.717	&	2.684	&	2.614	&	2.623	&	2.609	&	2.711	&	2.673	&	\\	
Indole-benzene (C$_s$ )		&	5.627	&	5.626	&	5.566	&	5.552	&	5.595	&	5.661	&	5.521	&	\\	
Phenol dimer (C$_1$ )		&	7.097	&	6.825	&	6.761	&	6.756	&	6.723	&	6.706	&	6.475	&	\\ [0.1 cm]
ME				&	$-$	&	0.09	&	0.06	&{\bf	0.02}	&	0.04	&	0.08	&	-0.03	&	\\
MAE				&	$-$	&	0.19	&	0.21	&{\bf	0.17}	&	0.22	&	0.20	&	0.18	&	\\ [0.1 cm]
\hline
\hline
ME				&	$-$	&	-0.01	&	-0.01	&	-0.12	&{\bf	-0.00}	&	0.07	&	-0.03	&	\\	
MAE				&	$-$	&	0.25	&	0.26	&	0.25	&	0.20	&{\bf	0.19}	&	0.22	&	\\	
\hline
\end{tabular}
\end{adjustbox}
\end{table*}

\section{Results}
All the calculations are done with the developer version of Q-CHEM simulation package~\cite{QCHEM2006}. 
For XC integrals the 99 points radial grid and 590 points angular Lebedev grid are used. Note that the non-bonded 
systems binding energies are very sensitive on the choice of the grid. The present choice of the grid is adequate and 
highly recommended for the complete energy convergence of the non-bonded systems~\cite{goerigk2017look,SCAND32016}. 
The test sets used in our calculations and the corresponding basis sets are mentioned 
in Table~\ref{test_set}. All calculations are performed with def2-QZVP basis set except the AHB21, IL16, WATER27,
DSCONF, and MG8 test sets, where the calculations are performed with the def2-QZVPD basis set. It is shown that the use of diffuse 
basis set drastically improves the results for those test sets~\cite{goerigk2017look}. 

\subsection{S22 test set}

To start with, we consider the S$22$ test set. As mentioned before it contains important non-covalent interacting 
molecules, that are often used for the benchmark calculations. The details of the different functional performance for the 
individual molecules are presented in Table~\ref{S22-full}. For reference values those obtained from CCSD(T)/CBS calculations by  
Sherrill et. al.~\cite{MarshallJCP2011} are considered. Regarding the performance of individual dispersion corrected 
functionals, we observe that all functionals perform in a impressive way. Regarding the H-bonded molecules, which consist of different complexes having 
biological interests, the NH$_3$ and H$_2$O dimer energies are overestimated by the -D3(0) and -D3(BJ) dispersion corrections, 
while for -D3(OP) the overestimation tendency is less evident. For other H-bonded systems, we also observe same 
tendency as -D3(OP), indicating its balanced performance for H-bonded systems.

\begingroup
\begin{table}
\scriptsize
\centering
\caption{ME and MAE (in kcal/mol) of different functionals for the S22 data set.
The best values are marked with bold style.}
\label{s22_comp1}
\begin{tabular}{lccccccccccccccccccccccccccccccc}\hline\hline

Methods		&ME	&	MAE\\	 \hline
\multicolumn{3}{c}{\bf semilocal/hybrid}\\[0.2 cm]
PBE$^a$             &-2.55  &2.55   \\
TPSS$^a$            &-3.44  &3.44   \\
SCAN$^a$		&-0.57	&0.91	\\
TM$^b$		&-0.53	&0.61	\\
revTM$^b$		&-1.80	&1.82	\\
M06-L$^a$		&-0.77	&0.81	\\
B3LYP$^a$           &-3.78  &3.78   \\
PBE0$^a$		&-2.33	&2.37	\\
TPSS0$^a$		&-3.06	&3.06	\\
\hline
\multicolumn{3}{c}{\bf semilocal+dispersion}	\\[0.2 cm]
PW86R-VV10$^d$	&	0.27	&	0.35	\\
$r$VV10$^c$	&	0.16	&	0.30	\\
SCAN+$r$VV10$^c$&	0.22	&	0.43	\\
SCAN+D3$^a$		&	0.38	&	0.45	\\
SCAN+D3(BJ)$^a$	&	0.45	&	0.42	\\
TM+D3(0)$^b$		&	-0.01	&	0.25	\\
TM+D3(BJ)$^b$	&	-0.01	&	0.26	\\
TM+D3(OP)$^b$	&	-0.12	&	0.25	\\
M06-L+D3(0)$^a$		&	0.44	&	0.52	\\
MS2+D3(OP)$^e$      &	$-$	&	0.43	\\
revTM+D3(0)$^b$	&{\bf	-0.00}	&	0.20	\\
revTM+D3(BJ)$^b$	&	0.07	&{\bf	0.19}	\\
revTM+D3(OP)$^b$	&	-0.03	&	0.22	\\ \hline
\multicolumn{3}{c}{\bf (range-separated)hybrid+dispersion}\\[0.2 cm]
B3LYP+D3$^a$	&       0.18	&	0.37	\\
B3LYP+D3(BJ)$^a$	&       0.29	&	0.31	\\
PBE0+D3$^a$		&	0.30	&	0.58	\\
PBE0+D3(BJ)$^a$	&	0.30	&	0.48	\\
revPBE0-D3(OP)$^a$  &       $-$	&       0.39	\\
TPSS0+D3$^a$	&	0.22	&	0.46	\\
TPSS0+D3(BJ)$^a$	&	0.19	&	0.38	\\
$\omega$B97X-D3$^a$	&	0.07	&	0.21	\\
$\omega$B97X-V$^a$	&	-0.10	&	0.22	\\
 \hline\hline
\label{table_s22}
\end{tabular}
\begin{flushleft}
 a-Ref. \cite{goerigk2017look}\\
 b-present work\\
 c-Ref. \cite{SCANVV102016} \\
 d-Ref. \cite{PhysRevB.86.165109} \\
 e-Ref. \cite{GordonD3OP2017} \\
\end{flushleft}
\end{table}
\endgroup

In case of the dispersion bonded systems, we observe a systematic slight underestimation of -D3(OP) functionals compared to
the -D3(0) and -D3(BJ) ones. Overall both the -D3(OP) corrected functionals underestimate the interaction energies. 

Next, for the mixed interaction, we observe underestimation or overestimation in the interaction energies from -D3(OP)
functional based on the interaction strength. Note that for this case the -D3(OP) balances more the interaction energies for 
individual molecules compared to the other two dispersion interactions.

To complete our analysis, in Table~\ref{s22_comp1}, we compare the ME and MAE of several popular GGA, meta-GGA and 
hybrid density functionals (global and range-separated). The dispersion corrected functionals are consistently 
improving their performance compared to the corresponding bare functionals. Note that revTM+D3(BJ) achieves the 
the best accuracy among the dispersion corrected semilocal functionals with MAE=0.19 kcal/mol, being significantly 
better than other dispersion corrected semilocal functionals. Within hybrid functionals, the $\omega$B97X-D3 is close to that of the revTM+D3(BJ).

\subsection{L7 test set}
The L7 test set consists of large sized complexes having dispersion dominated non-covalent bonds. Due to the 
computational efficiency, dispersion corrected semilocal XC functionals are very promising in case of such large 
complexes. Now, to test the accuracy of the above discussed methods, we apply both bare semilocal, and D$3$ 
corrected semilocal functionals to the optimized structures (TPSS-D/TZVP) of the complexes present in the L7 
test set\cite{L7JCTC2013}. This data set includes mixed hydrogen bonded complexes along with aliphatic, and 
strong aromatic dispersion bonded complexes. The binding energies of all the seven large complexes are shown 
in Table \ref{table_l7} considering all D3 corrected functionals and the CCSD(T) reference data \cite{L7CCSDT2018}. 
Among all the six dispersion corrected methods, revTM+D3(OP) has the least error with more accurately description 
of aromatic dispersion interactions(C3A, C3GC, C2C2PD) and hydrogen bonds (PHE). However, all the methods 
underbind the stacked Watson-Crick H-bonded guanine-cytosine dimer (GCGC) significantly. Such underestimation 
by TM based functionals is also reported in literature\cite{patra2019performance}. We also show the errors 
excluding the GCGC base pair from L7 data set in the lower panel of Table \ref{table_l7}. A drastic drop of
the MAE for all the cases can be seen and the revTM+D3(OP) is the best method with MAE=0.86 kcal/mol. 
Now, it is necessary to compare our methods with contemporary dispersion corrected methods to understand the 
hierarchy of development. So, we list the errors of L7 data set for above discussed methods along with errors of some 
available functionals in Table \ref{error_l7}. The TPSS+D3 method is proved to be best having least MAE value of 1.1 kcal/mol.
Note that the S30L benchmark set proposed in ref.~\cite{sure2015comprehensive} is more realistic than L7. We will consider 
these test cases in our future study.

\begin{table*}
\caption{Interaction energies (in kcal/mol) of L7 data set. The CCSD(T) reference values \cite{L7CCSDT2018} are given 
in the first column. The best values are marked with bold style.}
\label{table_l7}
\begin{adjustbox}{max width=\textwidth}
\begin{tabular}{lllccccccccccccccccccccccccccccccc}\hline\hline
L7 Complexes					&	CCSD(T)&	TM+D3(0)&	TM+D3(BJ)&	TM+D3(OP)&	revTM+D3(0)&	revTM+D3(BJ)&	revTM+D3(OP)&	\\ \hline
Octadecane dimer (CBH)				&	-11.6	&	-11.33	&	-10.09	&	-10.40	&	-12.08	&	-10.23	&	-10.91	&	\\
Guanine trimer (GGG)				&	-1.9	&	-2.08	&	-1.87	&	-1.85	&	-2.09	&	-1.71	&	-2.22	&	\\
Circumcoronene-Adenine dimer (C3A)		&	-17.0	&	-14.20	&	-13.86	&	-14.14	&	-14.71	&	-14.74	&	-15.81	&	\\
Circumcoronene-Guanine-cytosine dimer (C3GC)	&	-29.1	&	-25.12	&	-24.50	&	-24.76	&	-25.57	&	-25.58	&	-27.43	&	\\
Phenylalanine trimer (PHE)			&	-23.0	&	-24.90	&	-24.26	&	-24.58	&	-24.72	&	-24.27	&	-24.19	&	\\
Coronene dimer (C2C2PD)				&	-21.2	&	-16.70	&	-16.73	&	-17.02	&	-17.73	&	-18.66	&	-21.05	&	\\
Guanine-cytosine dimer (GCGC)			&	-12.8	&	-3.73	&	-3.15	&	-2.97	&	-2.95	&	-2.00	&	-3.14	&	\\ \hline
ME						&	$-$	&	-2.64	&	-3.15	&	-2.98	&	-2.39	&	-2.77	&{\bf	-1.69}	&	\\
MAE						&	$-$	&	 3.24	&	 3.52	&	 3.43	&	 3.07	&	 3.13	&{\bf	 2.12}	&	\\
								\multicolumn{8}{c}{Errors for L6 (removing GCGC from L7)}						\\
						&		&		&		&		&		&		&		&	\\
ME						&	$-$	&	-1.59	&	-2.07	&	-1.84	&	-1.15	&	-1.43	&{\bf	-0.36}	&	\\
MAE						&	$-$	&	2.27	&	2.49	&	2.36	&	1.94	&	1.85	&{\bf	0.86}	&	\\ \hline
\end{tabular}
\end{adjustbox}
\end{table*}

\begingroup
\begin{table}
\scriptsize
\centering
\caption{The ME and MAE (in kcal/mol) of different functionals for the L7 data set.}
\label{s22_comp}
\begin{tabular}{lccccccccccccccccccccccccccccccc}\hline\hline
Methods		&ME	&	MAE\\	 \hline
M06-L$^a$	&-3.0	&3.0	\\
M062X$^b$	&-3.2	&3.3	\\
SCAN$^a$	&-7.9	&7.9	\\
TM		&-8.0	&8.0	\\
revTM		&-15.0	&15.0	\\
\hline
PBE+D3$^a$	&-2.1	&2.6	\\
BLYP+D3$^b$	&2.1	&2.1	\\
TPSS+D3$^a$	&-0.9	&{\bf 1.1}	\\
SCAN+D3$^a$	&-1.2	&2.5	\\
M062X-D3$^b$	&{\bf -0.1}	&1.3	\\
TM+D3(0)	&-2.6	&3.2	\\
TM+D3(BJ)	&-3.1	&3.5	\\
TM+D3(OP)	&-2.9	&3.4	\\
revTM+D3(0)	&-2.3	&3.0	\\
revTM+D3(BJ)	&-2.7	&3.1	\\
revTM+D3(OP)	&-1.6	&2.1	\\
PBE0+D3$^a$	&1.4	&1.6	\\
B3LYP+D3$^b$	&1.7	&1.7	\\

 \hline\hline
\label{error_l7}
\end{tabular}
\begin{flushleft}
 a-Ref. \cite{SCANVV102016}\\
 b-Ref. \cite{PhysRevB.86.165109}\\
\end{flushleft}
\end{table}
\endgroup 

\subsection{Inter and intra-molecular non-covalent interactions}

\begingroup
\begin{table*}
\centering
\scriptsize
\caption{Mean errors and mean absolute errors (in kcal/mol) for benchmark test sets, using the D3-corrected semilocal 
XC functionals. For a better evaluation, we also provide 
the best semilocal+D3 and overall results for each test, taken from ref.~\cite{goerigk2017look}. The best values within TM and revTM based 
dispersion methods are marked with bold style.}
\begin{adjustbox}{max width=\textwidth}
\begin{tabular}{lcccccccccccccccccccccccccccccccccccccccccccccccccccc}\hline\hline
Test sets&	Errors	&		TM	&	TM	&TM	&	revTM	&	revTM	&revTM		&Best 		&Best  	\\	
	 &		&		+D3(0)	&	+D3(BJ)	&+D3(OP)&	+D3(0)	&	+D3(BJ)	&+D3(OP)	&Semilocal+D3	&Overall \\	\hline 
\multicolumn{10}{c}{Intermolecular non-covalent interactions (kcal/mol)} 									\\ [0.2 cm]
RG18	&	ME	&		0.01	&	0.01	&-0.02	&	-0.10	&	-0.14	&-0.13					\\
	&	MAE	&		0.21	&	0.19	&0.19	&	{\bf 0.15}&	0.18	&{\bf 0.15}	&0.09 		& 0.06 \\
        &               &                       &               &       &               &       	&		& (revPBE-D3(BJ))& (revTPSSh-D3(BJ))\\[0.2 cm]
ADIM6	&	ME	&		0.65	&	0.49	&0.43	&	0.36	&	-0.06	&-0.13				\\
	&	MAE	&		0.65	&	0.49	&0.43	&	0.36	&{\bf	0.06}	&0.13		&0.06 & 0.05 \\
	&       	&               	&               &       &       	&               &       	&(OLYP-D3(BJ))  &(BHLYP-D3(BJ)) \\[0.2 cm]
S22	&	ME	&		-0.01	&	-0.01	&-0.12	&	-0.00	&	0.07	&-0.03		&&&\\
	&	MAE	&		0.25	&	0.26	&0.25	&	0.20	&{\bf	0.19}	&0.22		&0.25         &0.14            \\
	&       	&               	&               &       &       	&               &       	&(BLYP-D3(BJ))  &(B2GPPLYP-D3(BJ)) \\[0.2 cm]
S66	&	ME	&		0.26	&	0.16	&0.09	&	0.16	&	1.10	&0.01							\\
	&	MAE	&		0.32	&	0.24	&0.21	&	0.22	&	0.16	&{\bf 0.15}	&0.17 &0.12 \\
	&       	&               	&               &       &       	&               &       	&(BLYP-D3(BJ))  &($\omega$B97X-V) \\[0.2 cm]
HEAVY28	&	ME	&		-0.07	&	-0.05	&-0.08	&	-0.19	&	0.15	&0.08							\\
	&	MAE	&		0.25	&	0.24	&0.25	&	0.32	&	0.22	&{\bf 0.18}	&0.23 &0.12 \\
	&       	&               	&               &       &       	&               &       	&(OLYP-D3(BJ))  &(MPW2PLYP-D3(BJ)) \\[0.2 cm]
CARBHB12&	ME	&		0.70	&	0.70	&0.65	&	0.90	&	1.01	&0.86							\\
	&	MAE	&		0.70	&	0.70	&{\bf 0.65}&	0.90	&	1.01	&0.86		& 0.44 & 0.22 \\
	&       	&               	&               &       &       	&               &       	&(M06-L-D3(0))  &(DSD-PBEB95-D3(BJ)) \\[0.2 cm]
PNICO23	&	ME	&		1.01	&	1.06	&0.99	&	0.85	&	1.31	&1.02							\\
	&	MAE	&		1.05	&	1.09	&1.03	&	{\bf0.94}&	1.32	&1.02		&0.25 &0.14 \\
	&       	&               	&               &       &       	&               &       	&(MN12L-D3(BJ))  &(PWPB95-D3(BJ)) \\[0.2 cm]
HAL59	&	ME	&		1.06	&	1.08	&1.05	&	0.87	&	1.29	&1.00							\\
	&	MAE	&		1.14	&	1.16	&1.14	&	{\bf 1.10}&	1.36	&1.12		&0.49 &0.29 \\
	&       	&               	&               &       &       	&               &       	&(M06-L-D3(0))  &(BHLYP-D3(BJ)) \\[0.2 cm]
AHB21	&	ME	&		-1.50	&	0.03	&0.08	&	-1.52	&	-0.10	&0.12							\\
	&	MAE	&		1.50	&	0.68	&{\bf 0.66}&	1.52	&	0.73	&{\bf 0.66}		& 0.47 &0.20 \\
	&       	&               	&               &       &       	&               &       	&(revTPSS-D3(BJ))  &(DSD-PBEB95-D3(BJ)) \\[0.2 cm]
CHB6	&	ME	&		-0.76	&	-0.79	&-0.77	&	0.09	&	-0.3	&-0.06							\\
	&	MAE	&		0.85	&	0.86	&0.85	&	0.53	&	0.66	&{\bf 0.50}	&0.45 &0.32 \\
	&       	&               	&               &       &       	&               &       	&(SCAN-D3(BJ))  &(MN15-D3(BJ)) \\[0.2 cm]
IL16	&	ME	&		-1.66	&	-0.36	&-0.25	&	-1.35	&	-0.28	&0.18							\\
	&	MAE	&		1.66	&	0.50	&0.46	&	1.35	&	0.42	&{\bf 0.38}	&0.31 &0.23 \\
	&       	&               	&               &       &       	&               &       	&(rPW86PBE-D3(BJ))  &(DSD-PBEP86-D3(BJ)) \\[0.2 cm]
	&		&			&		&	&		&		&		&					\\
			\multicolumn{10}{c}{Intramolecular non-covalent interactions (kcal/mol)}								\\
	&		&			&		&	&		&		&		&					\\		
IDISP	&	ME	&		0.87	&	0.79	&0.84	&	1.42	&	0.89	&1.26							\\
	&	MAE	&		3.75	&	3.71	&3.44	&	1.89	&	2.40	&{\bf 1.65}	&  2.05& 1.02\\
        &               &                       &               &       &               &       	&		& (SCAN-D3(BJ))& (DSD-BLYP-D3(BJ))\\[0.2 cm]
ICONF	&	ME	&		0.01	&	0.01	&0.01	&	0.07	&	0.05	&0.01							\\
	&	MAE	&		0.33	&	0.33	&0.34	&	0.30	&	0.29	&{\bf 0.26}	& 0.19&  0.14\\
        &               &                       &               &       &               &       	&		& (TPSS-D3(BJ))& (DSD-PBEP86-D3(BJ))\\[0.2 cm]
ACONF	&	ME	&		-0.22	&	-0.22	&-0.16	&	-0.00	&	-0.10	&-0.03							\\
	&	MAE	&		0.22	&	0.22	&0.16	&	{\bf 0.02}&	0.10	&0.04		&0.04 &0.03\\
        &               &                       &               &       &               &       	&		& (OLYP-D3(BJ))& ($\omega$B97X-V)\\
        &               &                       &               &       &               &       	&		&              & (revTPSS0-D3(BJ))\\[0.2 cm]
Amino20x4&	ME	&		0.11	&	0.11	&0.11	&	0.06	&	0.10	&0.08							\\
	&	MAE	&		0.25	&	0.27	&0.24	&	0.22	&	0.22	&{\bf 0.21}	&0.22&0.13\\
        &               &                       &               &       &               &       	&		& (SCAN-D3(BJ))& (B2GPPLYP-D3(BJ))\\
        &               &                       &               &       &               &       	&		&              & (DSD-BLYP-D3(BJ))\\[0.2 cm]
PCONF21	&	ME	&		-0.05	&	-0.03	&-0.08	&	-0.20	&	-0.21	&-0.06							\\
	&	MAE	&		0.66	&	0.65	&0.57	&	0.47	&	0.49	&{\bf 0.46}	& 0.47& 0.23	\\
        &               &                       &               &       &               &       	&		& (SCAN-D3(BJ))& (DSD-BLYP-D3(BJ))\\[0.2 cm]
MCONF	&	ME	&		0.43	&	0.38	&0.37	&	0.25	&	0.18	&0.14							\\
	&	MAE	&		0.56	&	0.53	&0.52	&	0.40	&	0.45	&{\bf 0.40} 	&0.33&  0.10	\\
        &               &                       &               &       &               &       	&		& (XLYP-D3(BJ))& (MPW2PLYP-D3(BJ))\\[0.2 cm]
SCONF	&	ME	&		0.46	&	0.46	&0.44	&	0.35	&	0.51	&0.35							\\
	&	MAE	&		0.74	&	0.76	&0.69	&	0.70	&	0.91	&{\bf 0.64} 	& 0.35& 0.06	\\
        &               &                       &               &       &               &       	&		& (M11L-D3(0))& (DSD-BLYP-D3(BJ))\\[0.2 cm]
UPU23	&	ME	&		-0.40	&	-0.32	&-0.37	&	-0.08	&	-0.01	&-0.02							\\
	&	MAE	&		0.49	&	0.48	&0.48	&	0.38	&{\bf	0.37}	&0.40		& 0.33&  0.33	\\
        &               &                       &               &       &               &       	&		& (revTPSS-D3(BJ))& (revTPSS-D3(BJ))\\[0.2 cm]
BUT14DIOL&	ME	&		-0.12	&	-0.12	&-0.16	&	-0.00	&	0.05	&0.01							\\
	&	MAE	&		0.17	&	0.18	&0.20	&	0.15	&	0.16	&{\bf 0.12}	&  0.18& 0.04\\
        &               &                       &               &       &               &       	&		& (revTPSS-D3(BJ))& ($\omega$B97X-V)\\[0.2 cm] \hline

TME&&0.04	&	0.17	&	0.15	&	0.10	&	0.28	&	0.23	&$-$&$-$\\
TMAE&&0.79	&	0.68	&	0.64	&	0.61	&	0.59	&	{\textbf{0.48}}	&$-$&$-$\\
\hline\hline
        \end{tabular}
\end{adjustbox}
\label{gmtkn_table}
\end{table*}
\endgroup

\begingroup
\begin{table*}
\caption{Mean errors and mean absolute errors (in kcal/mol) for the WATER27 benchmark test set, using 
the studied semilocal functionals along with
their dispersion corrected counterparts. The best semilocal+D3 and overall results are taken from ref.~\cite{goerigk2017look}.
TM and revTM results are from ref.~\cite{jana2020accurate}.}
\begin{adjustbox}{max width=\textwidth}
\begin{tabular}{lccccccccccccccccccccccccccccccccccccccccccccccccccc}\hline\hline
	Errors	&	TM&revTM&	TM	&	TM&TM&	revTM&	revTM&revTM&M06-L-D3(0)&DSD-BLYP-D3(BJ)  	\\	
		&		&&+D3(0)	&	+D3(BJ)&+D3(OP)&	+D3(0)&	+D3(BJ)&+D3(OP)&& 	\\	\hline
	        ME		&1.32&-1.24        &2.71&2.38&1.94&3.21&2.88&1.21&&	\\
		MAE		&1.44&\textbf{1.31}&2.79&2.45&2.02& 3.42&3.12&1.47&1.11 &0.94 \\
\hline
\end{tabular}
\end{adjustbox}
\label{tab5}
\end{table*}
\endgroup

\begingroup
\begin{table*}
\caption{Mean errors and mean absolute errors (in kcal/mol) for the in relative energies of the 
DSCONF Set of Conformers benchmark test set.  L1. M1 and S1 are the energetically most stable  
conformers for Lactose, Maltose, and Sucrose. The relative energies are calculated with respect to those 
stable conformers. The errors are calculated considering total 30 conformers. Best semilocal and double hybrid
functional results are also supplied from ref.~\cite{chan2020aqueous}.}
\begin{adjustbox}{max width=\textwidth}
\begin{tabular}{lccccccccccccccccccccccccccccccccccccccccccccccccccc}\hline\hline
	Errors	&	TM      &revTM&	TM+D3(0)	&	TM+D3(BJ)&TM+D3(OP)&	revTM+D3(0)&	revTM+D3(BJ)&revTM+D3(OP)&B-P86~\cite{chan2020aqueous}&DSD-PBE-P86~\cite{chan2020aqueous}\\	\hline
ME	&0.15	&	-0.24	&	0.23	&	0.23	&	0.20	&	0.12	&	0.19	&	-0.02	&		&		\\
MAE	&0.93	&	\textbf{0.69}	&	1.00	&	0.99	&	0.95	&	1.05	&	1.06	&	0.95	&	0.88	&	0.50	\\

\hline
\end{tabular}
\end{adjustbox}
\label{tab9}
\end{table*}
\endgroup


\begingroup
\begin{table*}
\caption{Mean absolute errors (in kcal/mol) for MG8 test set as calculated using different methods. The details of the test set 
and reference values are provided in ref.~\cite{chan2018formulation}.}
\begin{adjustbox}{max width=\textwidth}
\begin{tabular}{lccccccccccccccccccccccccccccccccccccccccccccccccccc}\hline\hline
	Group~\cite{chan2018formulation}&description~\cite{chan2018formulation}	&	TM      &revTM&	TM+D3(0)	&	TM+D3(BJ)&TM+D3(OP)&	revTM+D3(0)&	revTM+D3(BJ)&revTM+D3(OP)  	\\	\hline
NCED& noncovalent interaction (easy, cluster)&0.3&1.2&0.2&0.2&0.2&0.1&0.1&0.1\\
NCEC&noncovalent interaction (easy, dimer)&3.5&15.1&3.0&1.3&0.1&4.4&2.3&4.3\\
NCD&noncovalent interaction(difficult)&2.9&2.6&2.9&2.9&2.9&2.7&3.1&2.7\\
IE& isomerization energy (easy)&0.4&0.4&0.4&0.4&0.4&0.4&0.4&0.3\\
ID&isomerization energy (difficult)&18.0&17.5&17.5&17.7&17.4&17.0&17.1&17.6\\
TCE& thermochemistry (easy)&6.9&6.1&6.9&6.8&6.9&5.9&5.8&6.0\\
TCD&thermochemistry (difficult)&14.3&14.1&14.4&14.4&14.4&14.1&14.8&14.2\\
BH&barrier height&7.7&7.7&7.7&7.7&7.7&7.8&7.9&7.7\\
\hline
&MGCDB82&2.3&2.7&2.2&2.2&2.2&2.1&2.1&2.1\\
\hline
\end{tabular}
\end{adjustbox}
\label{tab10}
\end{table*}
\endgroup

\begin{figure}[ht]
 \begin{center}
 \includegraphics[width=3.0in,height=2.0in,angle=0.0]{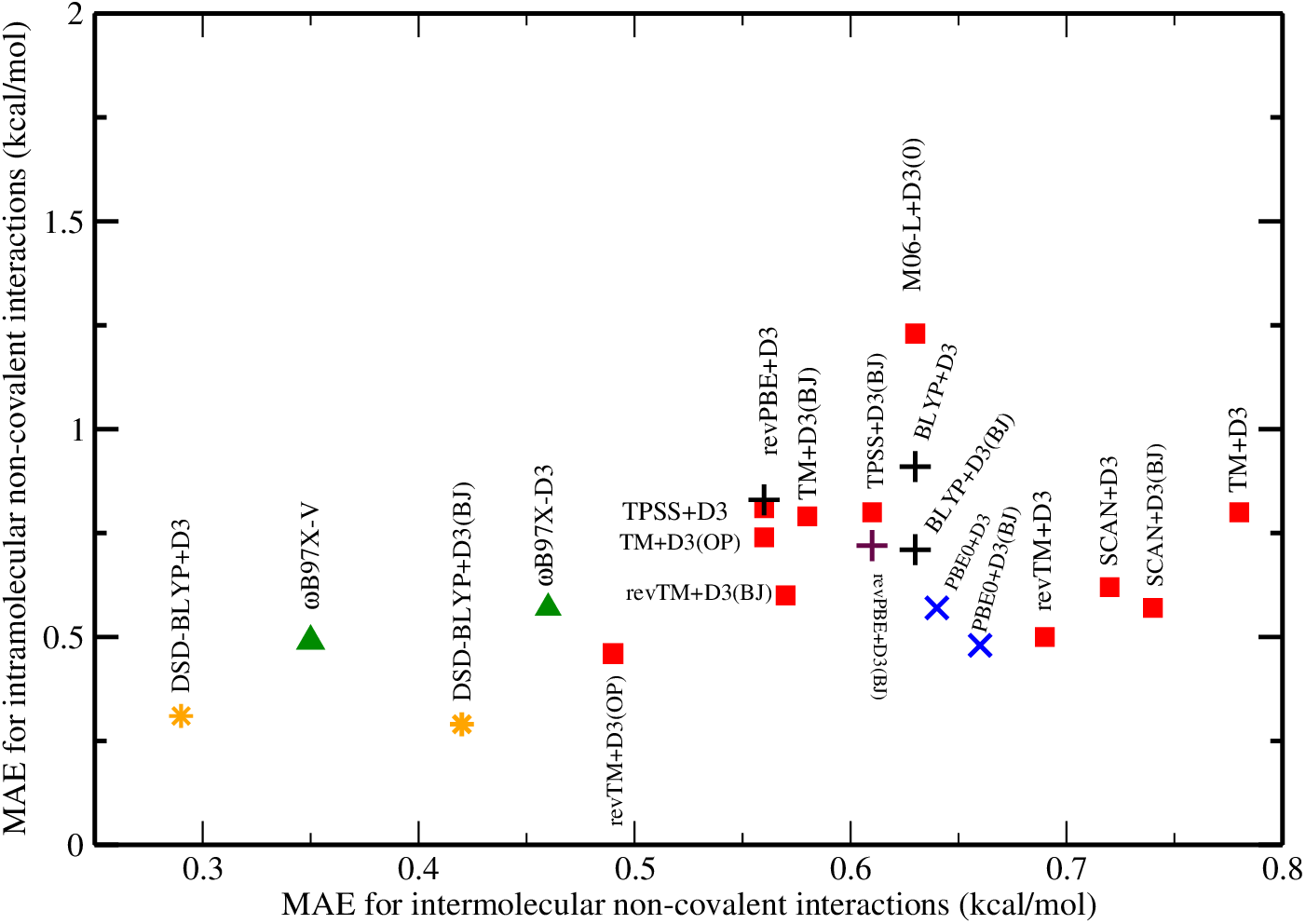}
  \end{center}
  \caption{Shown is the MAE (in kcal/mol) of inter-molecular non-covalent interactions versus the MAE 
(in kcal/mol) of intra-molecular
  non-covalent interactions for various functionals. Red-squares represent the meta-GGA+D3 functionals, 
black-crosses are GGA+D3 functionals, blue-x-shapes are global hybrids + D3, green-triangles are $vdW$-corrected 
long-range screened hybrids, and orange-stars are double hybrids. The WATER27 test set is not considered and the reference
error of the different funcionals (except TM and revTM based functionals) are taken from ref.~\cite{goerigk2017look}.}
  \label{figme}
\end{figure}

The inter-molecular binding energies of the dispersion bonded molecular complexes, arise from atoms of the 
two separate molecular systems. All the test sets and geometries are taken from the GMTKN55 database,  
where we do not include the WATER27, which is discussed separately within the hydrogen bonded complexes. 
Table~\ref{gmtkn_table} reports MAE of all the constructed dispersion corrected functionals, along with the
best dispersion corrected semilocal and the overall best method.

To start with, the RG18 test set contains the rare-gas dimers, trimers, tetramers, hexamers and complexes of rare gas 
with HF, ethyne, ethane and benzene. We obtain the best MAE from revTM+D3(OP) within the considered functionals with 
MAE=0.15 kcal/mol. In all cases, the -D3(OP) improves over -D3(0) and -D3(BJ) functionals. The ADIM6 test set consists of six alkane 
dimers binding energies. We observe revTM+D3(BJ) achieves the best accuracy among the semilocal D3 corrected functionals 
with MAE=0.06 kcal/mol, performing as the best semilocal-D3 result found from the OLYP-D3(BJ) functional. 
Therefore, for alkane dimers binding energies, revTM+D3(BJ) is quite a good candidate. Considering the S22 test set, it was 
already discussed in the previous section. In this case, revTM+D3(BJ) achieves the best accuracy among the semilocal D3 theory, better 
than so far best BLYP-D3(BJ). Similar accuracy is also observed for the S66 test set with the revTM+D3(BJ) functional. However, 
in this case we observe revTM+D3(OP) bit better than revTM+D3(BJ). This is due to the better performance of revTM+D3(OP) for 
H-bonded systems. The HEAVY28 test set consists of non-covalent binding energies of 28 heavy-element-hydride dimers. In this 
case also, revTM+D3(OP) outperforms other dispersion corrected functionals with MAE=0.18 kcal/mol. The CARBHB12 test set  
represents 12 hydrogen-bonded complexes of carbene bound with CClCH$_3$, SiH$_2$, and 1,3-dimethylimidazol-2-ylidene. Though 
the TM+D3(OP) gives MAE=0.65 kcal/mol, still M06-L-D3(0) is the best dispersion corrected semilocal functional with 
MAE=0.44 kcal/mol. For PNICO23 test set, all considered functionals overestimate the binding energies 
corresponding to the most accurate semilocal D3 approach MN12L-D3(BJ). The HAL59 test set represents non-covalent binding energies of 
halogenated dimers, being constructed from the combination of XB51~\cite{HAL59JCTC2013} and X40~\cite{HAL59JCTC2012} 
test sets. In this case also the slight overestimation is observed from all the dispersion corrected functionals,
and the lowest MAE of 1.10 kcal/mol is obtained from revTM+D3(0), while the best semilocal D3 corrected functional 
is the M06-L-D3(0) with MAE=0.49 kcal/mol. The AHB21 test set contains the interaction  energies of 21 anionic and 
neutral dimers. The TM+D3(OP) and revTM+D3(OP) are performing better compared to the others functionals, because -D3(OP)
performs in a more balanced way for H-bonded and dipole-interacting systems. The revTM+D3(OP) is also performing comprehensively
for the six cationic$-$neutral dimers test set CHB6. Next for the IL16 test set which consists of 16 cation$-$anion 
non-covalently bonded model dimers, revTM+D3(OP) performs quite well and very close to the most accurate semilocal D3
methods rPW86PBE-D3(BJ).

Next, we perform the assessment of the dispersion corrected functionals for the intra-molecular binding energies of 
the GMTKN55 database. It consists of dispersion interactions within the same molecular complex. To start with, we consider the IDISP12 
test set having interactions of large hydrocarbon molecules. The revTM+D3(OP) achieves the best accuracy in this case 
with MAE=1.65 kcal/mol and its accuracy is better than so far the best semilocal D3 method SCAN-D3(BJ). Similar accuracy of 
revTM+D3(OP) is also observed for the ICONF test set that consists of non-covalent interactions of inorganic molecules. The accuracy 
of the revTM+D3(OP) is also very prominent for the ACONF test set which contains relative energies of 15 $n$-butane, 
$n$-pentane and $n$-hexane conformers. It also gives the very similar accuracy as that of the so far best dispersion corrected 
semilocal method OLYP-D3(BJ). The revTM+D3(OP) is also very accurate for AMINO20$\times$4 complexes having amino acids as a 
base pair. The peptide bonds within the amino acid are crucial for bio-molecular systems, as DNA and RNA pairs. 
The extended PCONF21 test set consists of relative energies of eleven phenylalanyl-glycyl-glycine tripeptide and ten 
tetrapeptide conformers respectively. Also in this case, revTM+D3(OP) is surprisingly the most accurate method within various 
semilocal+D3 approximations with MAE=0.46 kcal/mol. The accuracy of the revTM+D3(OP) is more evident (with MAE 0.40 kcal/mol) 
from MCONF test set which consists of relative binding energies of 52 melatonin having quadrupole$-$dipole, aromatic$-$amide, and 
hydrogen bond interactions important for biomolecules. The SCONF test set consists of 14 and 3 relative energies of 
3,6-anhydro-4-O-methyl-D-galactitol and b-D-glucopyranose conformers, respectively. In this case also, revTM+D3(OP) 
performs better than the other dispersion corrected functionals, while the best semilocal+D3 method is the M11L-D3(0). For 
UPU23 test set, all dispersion corrected revTM perform with almost same accuracy. Same is true for the TM based 
dispersion corrected methods. This test set consists of nucleic acids and biomolecules which are the main 
constituents of RNA. Finally, for the BUT14DIOL test set, which consists of strong intra-molecular hydrogen bonds, 
the revTM+D3(OP) is the best within the semilocal+D3 methods.

To make our comparison of the accuracy of different popular functionals in a more competitive manner, in 
Fig.~\ref{figme}, we plot the MAE of the inter-molecular non-covalent interactions versus the MAE of intra-molecular
non-covalent interactions. It is noticed that, at the semilocal level, revTM+D3(OP) achieves the best accuracy.
Moreover, the revTM+D3(OP) functional is even better than the $\omega$B97X-V and $\omega$B97X-D for the intra-molecular 
non-covalent interactions, where both functionals are the range-separated hybrids and quite expensive for large 
molecular systems. Note also that revTM+D3(OP) is better than well known hybrid+D3 functionals like PBE0+D3(BJ) in both cases.


\subsection{Water clusters}

The remarkable accuracy of the -D3(OP) based semilocal fnctionals is also clearly evident from Table \ref{tab5}, 
where we assess the dispersion corrected semilocal functionals for various water clusters.
This test 
set includes H-bonded water clusters which are either neutral or positively, and negatively charged. This test set 
is extracted from the 
GMTKN55 database~\cite{goerigk2017look} as mentioned before, in order to emphasize the performance of the 
functionals for 
H-bond within water molecules. It is seemingly quite interesting that the -D3(OP) does not deteriorate the 
performance of TM and revTM functionals, unlike other -D3 methods. The bare TM and revTM give the MAE of about 1.44 
kcal/mol and 1.31 kcal/mol, respectively,
which are only slightly better than 2.02 kcal/mol and 1.47 kcal/mol obtained upon addition of 
the -D3(OP) correction. These results 
motivate us to further study the -D3(OP) corrected TM and revTM functionals for water properties. Note that very recently the 
revTM functional is assessed for different water properties~\cite{jana2020accurate} and found to be very accurate for 
different water properties. In this case 
M06-L-D3(0) is the best dispersion corrected semilocal functional with MAE=1.11 kcal/mol and overall DSD-BLYP-D3(BJ) 
is the best functional with MAE=0.94 kcal/mol. 


\subsection{Conformers for lactose, maltose, and sucrose}

Energetic of the bio-molecular conformers are important in various applications of chemical and biological systems. 
Being very large structures, the semilocal XC are the most preferred method to simulate those systems. Here, we studied 
relative energies of the different conformers of the lactose, maltose, and sucrose using the prescribed methods. This test set 
(DSCONF) is proposed recently~\cite{chan2020aqueous}. Note that the basic constituent of these conformers are the amino acids
and peptides having hydrogen bonds. Therefore, it is an interesting test case because a major factor of this test set is determined 
by the relative conformer energies of OH-O hydrogen bond, similar to the WATER27 test set. The error statistics as obtained 
from different functionals are reported in Table~\ref{tab9}. We observe the revTM becomes the most accurate functional with 
MAE=0.69 kcal/mol followed by the bare TM functional. Similar to the WATER27 test case the D3-0 and D3-BJ variants work less 
well than D3-OP.

\subsection{Small representative MG8 thermochemical test set}

Lastly, we assess the constructed functionals performance for the small representative MG8 thermochemical test set.
The MG8 test set is proposed recently~\cite{chan2018formulation} and it represents statistically accurate depiction of the MGCDB84
test set~\cite{mardirossian2017thirty}. This test set contains $64$ data points
instead of the large 5000 data points of the MGCDB82 test set. Like MGCDB84 on which it is based, MG8 divides the data into different 
types of properties like noncovalent interactions, isomerization energies, thermochemical properties, and barrier heights. The details of the 
test set and its benchmark values can be found in ref.~\cite{chan2018formulation}. The MAEs of the each test set as obtained form different 
functionals are listed in Table~\ref{tab10}. The MAEs for MGCDB82 are also calculated in Table~\ref{tab10} using the formula suggested in 
Eq.(1) of ref.~\cite{chan2018formulation}. It is obvious that the isomerization energy and thermochemistry of difficult cases are particularly challenging, though that is generally 
true for most functionals; for example, even B97M-V has an MAE over 10 kcal/mol for isomerization energy~\cite{chan2018formulation}. In this respect, the dispersion
corrected semilocal functionals show improvement in a systematic way than its bare functionals.  Interestingly, the performance of
the -OP corrected functionals is quite promising. 



\section{Conclusions}

We have assessed the benchmark calculations of the 
D3-corrected TM and revTM meta-GGA XC functionals, for 
a large palette of molecular complexes, characterized by various 
non-covalent interactions, such as inter- and intra-molecular 
dispersion, hydrogen, halogen, dihydrogen, dipole-dipole and mixed bonded systems.
We have constructed several forms of the D3-functionals, using 
the zero, rational damping, and optimized parameter damping functions. A total of six 
variants of dispersion corrected
functionals are tested for a wide range of interesting systems. 
Our primary focus has been to measure the accuracy and 
applicability of the proposed
methods for different kinds of dispersion interactions. It turns out that within the $vdW$-corrected semilocal 
approximations, the revTM+D3(OP) gives an outstanding performance, outclassing many popular functionals, and 
competing with the expensive dispersion corrected range-separated hybrids $\omega$B97X-D and $\omega$B97X-V. 

For the energetic of the non-covalent binding energies, the performance of revTM+D3(OP) is obtained to be very good 
for S22, L7, various
inter- and intra-molecular non-covalent interaction test sets of the well-known GMTKN55 database, and the 
H-bond interaction of charged moieties with neutral small molecules. For comparison purposes, we discuss 
separately the binding
energy of the neutral and charged water clusters, where the inclusion of the -D3 within semilocal functional usually 
over-binds the energies. Interestingly, we observe that 
TM+D3(OP) and revTM+D3(OP) do not deteriorate much the accuracy of the bare functionals. The impressive performance of 
(rev)TM and (rev)TM+D3(OP) is also more evident from the relative conformer energies is OH-O
hydrogen bond of the lactose, maltose, and sucrose. Overall, for the small representative MG8 thermochemical test set
also the ``OP'' corrected functionals performance in an impressive manner.


Overall, revTM-D3 XC functional delivers awe-inspiring performance and acquire excellent accuracy close to the 
computationally costly range-separated hybrids and double-hybrid functionals. Importantly, it performs well for 
different interaction ranges of the non-covalent systems and can be considered as an important dispersion corrected 
functional within the dispersion corrected density functional theory zoo. As a concluding remark it 
is also important to note that recently the -D4 dispersion correction of Grimme shows its productive power over -D3,
which we will consider in our future assessment.

\section{Acknowledgements}
A.P. would also like to acknowledge the financial support from the Department of Atomic Energy, Government 
of India. S.J. would like to thank Prof. Stefan Grimme for many useful and technical suggestions. S.J. would also like to thank Prof. 
Bun Chan for providing much useful technical information about the 
calculation of the DSCONF, MG8 test set.  This work has been performed in a 
high-performance computing facility of NISER. PS would like to thank  Q-Chem, Inc. and developers 
for providing the code.

\section{Data available on request from the authors}

The data that support the findings of this study are available from the corresponding author upon reasonable request.

\twocolumngrid
\bibliography{tmd3}
\bibliographystyle{apsrev4-1}

\end{document}